\documentclass{vgtc}                          

\ifpdf
  \pdfoutput=1\relax                   
  \usepackage{graphicx}                
  \DeclareGraphicsExtensions{.pdf,.png,.jpg,.jpeg} 
\else
  \usepackage{graphicx}                
  \DeclareGraphicsExtensions{.eps}     
\fi%

\graphicspath{{figures/}{pictures/}{images/}{./}} 

\usepackage{microtype}                 
\PassOptionsToPackage{warn}{textcomp}  
\usepackage{textcomp}                  
\usepackage{mathptmx}                  
\usepackage{times}                     
\usepackage{cite}                      
\usepackage{tabu}                      
\usepackage{booktabs}                  

\onlineid{0}

\vgtccategory{Research}

\vgtcinsertpkg

\title{Empowering People with Intellectual and Developmental Disabilities through Cognitively Accessible Visualizations}

\author{Keke Wu\thanks{e-mail: kekewu@cs.unc.edu}\\ %
        \scriptsize University of North Carolina at Chapel Hill %
\and Danielle Albers Szafir \thanks{e-mail: danielle.szafir@cs.unc.edu}\\ %
     {\scriptsize University of North Carolina at Chapel Hill}}

\abstract{Data has transformative potential to empower people with Intellectual and Developmental Disabilities (IDD). However, conventional data visualizations often rely on complex cognitive processes, and existing approaches for day-to-day analysis scenarios fail to consider neurodivergent capabilities, creating barriers for people with IDD to access data and leading to even further marginalization. We argue that visualizations could be an equalizer for people with IDD to participate in data-driven conversations. Drawing on preliminary research findings and our experiences working with people with IDD and their data, we introduce and expand on the concept of cognitively accessible visualizations, unpack its meaning and roles in increasing IDD individuals' access to data, and discuss two immediate research objectives. Specifically, we argue that cognitively accessible visualizations should support people with IDD in personal data storytelling for effective self-advocacy and self-expression, and balance novelty and familiarity in data design to accommodate cognitive diversity and promote inclusivity.%
} 

\CCScatlist{
  \CCScatTwelve{Human-centered computing}{Visualization}{}{};
  \CCScatTwelve{Human-centered computing}{Accessibility}{}{}
}

\begin{document}

\teaser{
  \centering
  \includegraphics[width=\linewidth]{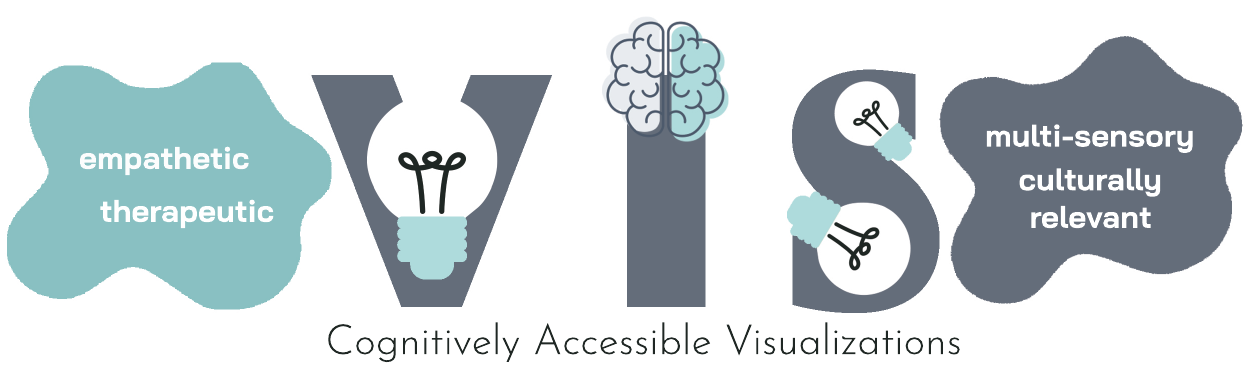}
  \caption{%
  	Two immediate objectives of cognitively accessible visualizations are to design for personal storytelling and to balance novelty and familiarity in data design. Particularly, the visualizations need to facilitate empathy in the viewer and support therapeutic expressions for people with IDD. In addition, designers may leverage multi-sensory approaches to representing data, and create culturally relevant interfaces to be more in tune with the needs of this community. %
  }
  \label{fig:teaser}
}


\maketitle

\section{Introduction} 
Data plays an important role in understanding, addressing, and advocating for the needs and rights of people with Intellectual and Developmental Disabilities (IDD) \cite{datachi}. Researchers, advocates, and organizations can use data to create interventions to improve lives of people with IDD \cite{data4health, personcentered, braddock}. Collecting and analyzing data about people with IDD provides insights into challenges faced by the community \cite{data4health}, helps identify disparities and inequalities in resource allocation \cite{personcentered}, and guides evidence-based policy-making and drives positive social change \cite{braddock}. Data can also empower individuals with IDD. It allows the monitoring of personal health, behaviours and progress, assists with autonomy and enables people with IDD to make informed decisions and develop effective advocacy strategies \cite{datachi}. However, working with data is complicated and requires various cognitive skills to effectively analyze, interpret, and draw insights \cite{chi}. People with IDD face significant limitations in many cognitive areas; such as learning, reasoning, abstract thinking, attention, concentration, memory and recall \cite{faqs}, which prevent them from effectively using, understanding, and making decisions with data, often excluding them from data analytics and leaving them vulnerable to many ethical issues. Still, little progress has been made to improve data accessibility for people with IDD: society tends to focus on their disabilities rather than abilities \cite{JansenvanVuuren2020StigmaAA}, underestimate their potential and trivialize their needs for data analytics. In addition, each individual with IDD can have a significantly different cognitive profile, which makes developing universal solutions particularly challenging. 

Since 2018, we have been working with this population to design cognitively accessible visualizations. Our investigation started from a collaboration with an initiative that aims to support people with IDD in data-driven financial self-advocacy \cite{braddock}. In past work, we conducted a graphical perception experiment to understand how various visualizations may impact data accessibility in the context of budgetary data analysis \cite{chi}. Overall, the study underlined the need to design visualizations that attend to the specific abilities and preferences of people with IDD and demonstrated that cognitively accessible visualizations may require a different set of guidelines. We then explored more broadly the lived data experiences of people with IDD through semi-structured interviews \cite{datachi}. We found that people with IDD frequently used personal data for self-advocacy, self-expression and everyday functioning. However, many of these data encounters were invisible and inaccessible to people with IDD, and were not well supported by current data visualizations, leaving them with limited access to data important to their well-being. Collectively, these studies suggest that visualizations have the potential to empower people with IDD and that visualization research needs to pay more attention to the specific needs of this population to develop truly accessible solutions.

In this article, we build on these findings and our own experiences to identify a critical near-term research agenda for increasing cognitive access to data. We argue that cognitively accessible visualizations could serve as an equalizer for people with IDD to participate in society. Specifically, we discuss two immediate research priorities: supporting people with IDD in personal data storytelling for effective self-advocacy and self-expression, and balancing novelty and familiarity in data design to accommodate cognitive diversity and promote inclusivity.

\section{Supporting Personal Storytelling with Data}
\begin{figure}[h]
\centering
\includegraphics[width=8cm]{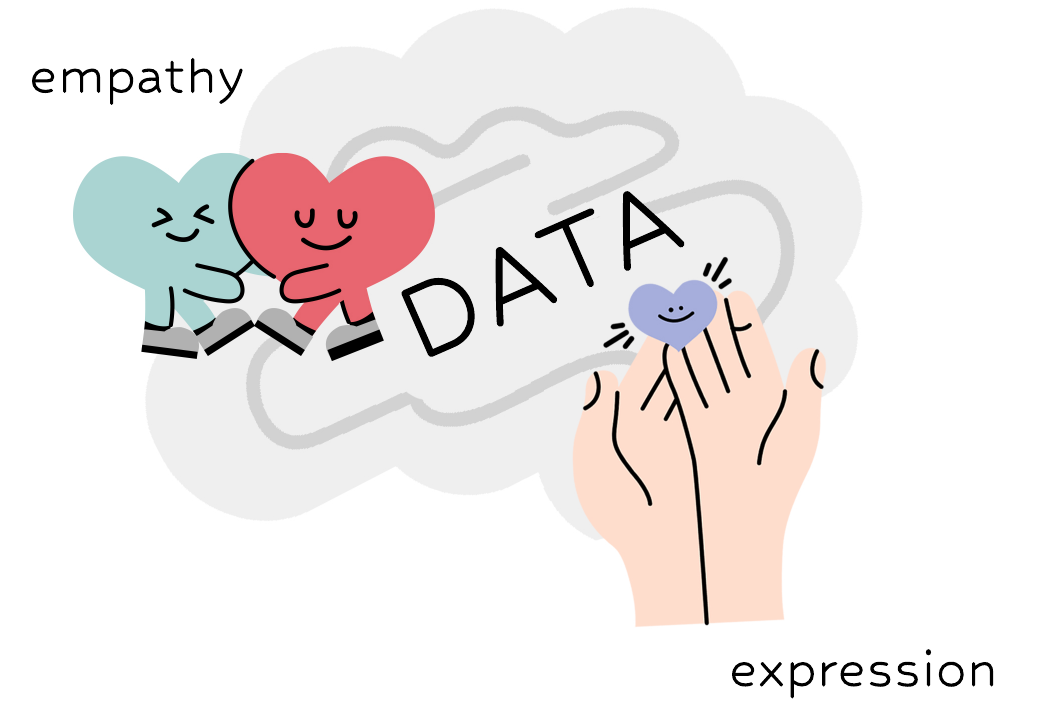}
  \caption{%
  	Cognitively accessible visualizations should support people with IDD in telling personal stories through data to elicit empathy in the viewer for more effective self-advocacy and facilitate therapeutic expression and self-compassion by helping them make sense of personal data and life experiences.%
  }
  \label{fig:storytelling}
\end{figure}

People with IDD use data in a variety of contexts, such as advocating for one's needs and rights, sharing personal experiences, or expressing thoughts and feelings through data as a creative outlet. However, our preliminary findings suggest that people assemble and visualize data sets to tell personal stories 
in a 
manual and ad hoc manner. With limited visualization literacy, most participants found traditional storytelling approaches inaccessible and often intimidating to consume and author, leaving them unable to leverage data to their full benefit in self-advocacy. Therefore, visualization research needs to understand how to support effective personal data storytelling for people with IDD (Fig. \ref{fig:storytelling}). Visualizations need to represent personal data of people with IDD to help build empathy through shared lived experience and turn this empathy into actions. Visualization tools should empower people with IDD to better communicate personal stories for creative expression, self-reflection, and compassion. 

\subsection{Designing for Empathetic Persuasion}
Visualization research should aim to develop best practices for designing data stories that build empathy in the viewers to help them better understand the lived experience of people with IDD. Many individuals with IDD experience stigma, prejudice and discrimination: they are largely viewed as liabilities to be managed rather than assets that contribute unique perspectives to their community\cite{JansenvanVuuren2020StigmaAA}. These individuals usually live in systems that were designed without their unique needs in mind and, as a result, this population often faces barriers to engagement and participation, ranging from the health-care system to the criminal justice system to employment and education \cite{ada}. Our research showed that data could be an empowering tool for self-advocacy and help shift these negative narratives for better social inclusion. Specifically, we found that people with IDD were trying to make compelling arguments through collecting and sharing data from their personal lives. These could be written logs of daily activities and accomplishments,  multimedia files that demonstrate competence and capability, or even survey results that illustrate the life expectancy gap between disabled and non-disabled populations.

To translate this data into powerful arguments, cognitively accessible visualizations should more vividly represent the stories of people with IDD; they could help communicate their life aspects, challenges, thoughts and feelings to those who may not share the experiences; and evoke empathy in the viewers. To design such compelling data visualizations, or to more effectively move a viewer to action, we recommend designers use storytelling techniques, but center the design of these stories around the particular person described by the data and/or the argument to make rather than the data \textit{per se}. This means that designers may need to co-design with people with IDD to build better empathy and an understanding of their lives, and then base the design on IDD individuals' point of view. Visualizations may help a viewer better relate to the story by using true-to-life imagery and having a character that carries the IDD individual's values and beliefs. To better elicit empathetic responses, future authoring tools may learn from principles of the arts to create an aesthetic experience with data. For example, different colors can be mapped to emotions and feelings; visual emphasis and contrast can be made to highlight certain piece of data or make a statement; music, sound effects and narrations can be layered into a multimedia data visualization to establish a setting, develop characters, or advance the plot. These aesthetic experiences may allow a viewer without IDD to feel the emotions and perspectives of people with IDD, being able to peek into their inner worlds and daily lives, getting a sense of ``us'' rather than ``them.'' Such stories may help people develop a strong sense of engagement \cite{ognjenovic1997psiholovska} and deeper connection with characters depicted in the story (i.e., people with IDD),  and activate empathy, which then translates to action. 

\subsection{Designing for Therapeutic Expression} 
Visualization can serve as a therapeutic tool for people with IDD to regulate emotions and develop self-compassion by encouraging creative expression and meaning-making with personal data. People with IDD are three to four times more likely to develop psychiatric disorders compared with the general population \cite{book}. Approximately one third of adults with ASD have emotion dysregulation and challenging behaviours (CBs) \cite{cbs}, negatively affecting their quality of life and social participation. We found that data was commonly used by people with IDD, especially those with ASD, as a way of self-expression and reflection, a proxy to communicate thoughts and feelings in psychiatric treatments, and occasionally an educational device to improve self-regulation \cite{datachi}. For example, participants kept written notes of their symptoms and reactions to social situations, they created a values checklist to reflect everyday behaviors, curated video clips that depict human interactions to discuss with their therapists, and created drawings, photos, and multimedia files to express how they feel about themselves and the world around them. 

People with IDD, particularly those with ASD, usually experience anxiety as a result of sensory processing issues \cite{sensory}. Our studies indicate that the anxiety may also stem from the social awareness that they are different: they may experience feelings of loneliness and isolation as a result of not conforming to social norms or an inability to fit in with their peers. For example, one participant described how they personally relate to the movie \textit{Beauty and the Beast}, \emph{``My understanding is that the beast is autistic to a certain extent. He has a lot of books. He loves knowledge. That's essentially me. And he's nice, but he has that inner world that's not being understood. And Belle is just this oddball. She just seems to be ahead of her time that she understands the beast. She's like the therapist that he needs and she ends up falling in love with the beast.''} Similarly, another participant from our workshop on the theme of ``Aliens from the VisuaLand'' expressed how they ``\textit{truly feel like an alien to the human world}.'' The sense of alienation can magnify feelings of self-doubt, resulting in social anxiety, low self-esteem, and depression, ultimately leading people with ASD to shut down. Due to the many shared behavioral, conceptual, communicative, and social difficulties, individuals with IDD in general experience high risks of co-morbid mental health conditions such as affective, anxiety, psychotic, and impulse control disorders \cite{mentalhealth}. However, research indicates a common lack of clinical knowledge and training about the needs of people with IDD among mental health professionals, resulting in misdiagnosis and mistreatment that are associated with an exacerbation of dysfunction and disability \cite{mentalhealth}. Non-pharmacological therapeutic support in IDD—such as social prescribing, behavioral and educational interventions, or psychotherapy—seem to be more effective than other interventions \cite{intervention}. While beyond traditional visual analytics, data visualizations do not have to be purely analytical. We argue that cognitively accessible visualizations could be a therapeutic device, providing a creative outlet for individuals with IDD to express themselves, communicate ideas, and make sense of complex life experiences for positive meaning-making. This can happen in several ways: 

\noindent\textbf{(1) Supporting self-expression and storytelling.} Data visualization can be a creative way to tell personal stories or narratives. By combining data and personal experiences, individuals with IDD can create visual representations that highlight their unique perspectives or journeys. By utilizing design choices such as colors, shapes, and visual metaphors, individuals can infuse their data visualizations with emotion and personal expression. This can help overcome communication challenges and enable people with IDD to create visualizations that evoke specific feelings, connect with the audience on a deeper level, and convey the subjective aspects of the data being represented. Visualization research may further investigate the connection between different design elements and emotions to provide templates for people with IDD to readily communicate feelings and generate stories from the data. 

\noindent\textbf{(2) Encouraging self-reflection and compassion.} Engaging in data visualization can prompt people with IDD to reflect on their own data, habits and behaviours. People with IDD can gain insights into their patterns, progress, or areas for improvement. Visualization can serve as a means to explore personal interests or passions. Whether it's visualizing personal fitness data, travel experiences, or reading habits, individuals can create visual representations of their hobbies, interests, or personal pursuits. This allows them to share their enthusiasm, express their unique perspectives, and potentially inspire others who resonate with their visualizations. Visualization research may explore different encoding strategies and representation formats for expressing these personal experiences to help people with IDD intelligently manage their data and arrive at useful insights.
 
\section{Balancing Novelty and Familiarity in Data Design} 
Despite being everywhere, data is invisible to many people with IDD, who often consider data as a remote subject and show apathy or even aversion to it. Most of them are usually unconscious of everyday data experiences and disinterested in improving data literacy \cite{datachi}. Moreover, data is abstract and usually comes in diverse formats in large volumes, which requires significant cognitive resources and specialized training to work with. Many participants in our studies complained about the complexity of data and their lack of abilities and resources to make sense of it \cite{datachi}. Evidence from our research shows that when designed carefully, visualizations are able to help people with IDD better reason with data \cite{chi}, and that the awareness of and willingness to work with data is largely driven by personal motivation \cite{datachi}. However, we found that most participants associate data with ``\textit{hard drive}'' and ``\textit{technology}'', treating it as something used by ``\textit{a computer type of gal}'' or ``\textit{people who do studies}'' but not themselves. These stereotype-driven beliefs about data can have detrimental effect on people's ability and confidence in taking control of data. To change this stereotypical view and encourage interests in working with data, cognitively accessible visualization should strike for a balance of novelty and familiarity in design. We anticipate that this can best be achieved by creating novel multisensory experiences to increase the awareness and appeal of ``mundane data'', and also by having IDD individuals' shared knowledge and experiences in mind to design culturally relevant interfaces that help them digest and relate to new information.

\subsection{Creating Multisensory Data Experiences}
Visualization should accommodate the diverse needs and abilities of people with IDD by providing alternative representation options and novel data experiences. On top of being highly susceptible to psychiatric disorders, individuals with IDD are also far more likely to develop multiple physical health conditions, such as epilepsy, visual impairment, hearing loss, mobility issues, and sensory processing disorders \cite{physical, healthproblem, 10.1371/journal.pone.0256294}. Our data interview highlighted a common yet unconscious use of data among this population, especially those less independent participants: tracking personal routine data to adjust daily regimen \cite{datachi}. However, this data is usually invisible and inaccessible to people with IDD, as it is considered as part of the daily routine and is collected for other people (e.g., caregivers, psychologists, dietitians, physicians etc.) rather than themselves, which significantly reduces the agency and control of their own data and even bodies. We argue that cognitively accessible visualizations should help IDD individuals improve awareness of personal data, encouraging them to explore this data and participate in decision-making about their own lives. 

\begin{figure}[h]
\centering
\includegraphics[width=4.2cm]{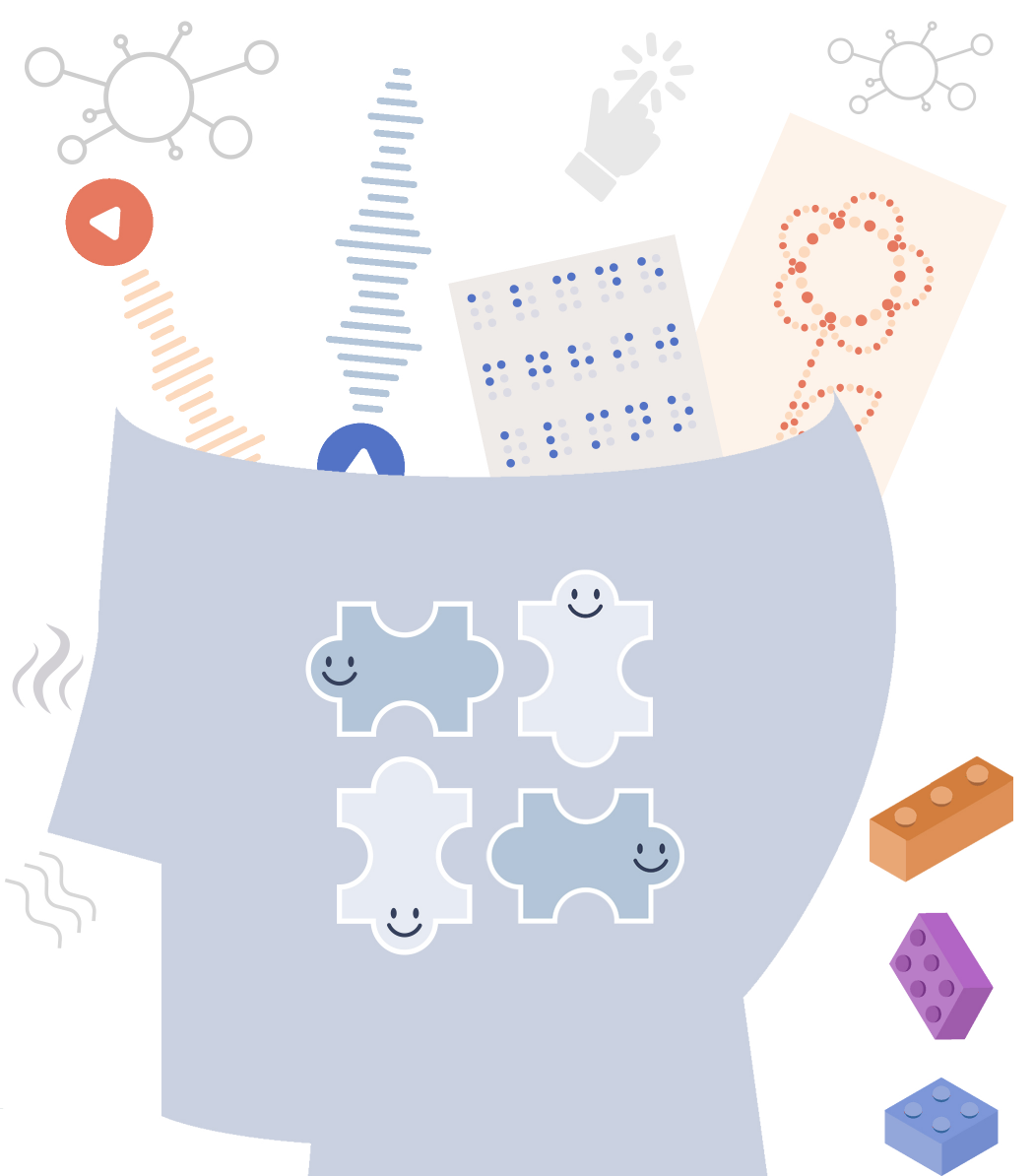}
  \caption{%
  	Cognitively accessible visualizations may take in multisensory forms for better data awareness and accessibility.%
  }
  \label{fig:sensory}
\end{figure}

Discussions with caregivers and therapists suggested that individuals with IDD will likely benefit from a combination of visual, verbal, and tactile cues to understand prompts and complete everyday tasks. Given the heterogeneity of this population, we suggest visualization designers consider using multisensory data representation strategies to help people with IDD reason with data \ref{fig:sensory}. For example, data can be transformed into \textbf{physical sculptures or installations} to provide a tactile and visual experiences \cite{datasculpture}. Individuals with IDD can physically interact with the sculpture, manipulating different elements to explore and reason with the data through a constructionist approach \cite{constructive, DIgnazio2018CreativeDL}. The sculpture could use various textures, shapes, or colors to represent different data points or categories, allowing individuals to physically rearrange and compare them \cite{DRS}. Data can also be \textbf{sonified}, with different data attributes being mapped to various auditory elements such as pitch, rhythm, or volume. This allows individuals to listen to patterns, trends, or relationships in the data, providing an alternative perspective for reasoning and analysis 
 \cite{10.1111:cgf.14298}. Utilizing \textbf{tactile interfaces}, such as touchscreens, haptic feedback devices, or braille displays, can enable individuals to reason with personal data through touch. People can feel patterns, textures, or vibrations associated with different data points, enabling a tactile exploration and analysis of the data \cite{10.1111:cgf.14298}. By turning data exploration and analysis into \textbf{interactive games or challenges}, individuals can utilize visual, auditory, and even kinesthetic senses to make decisions, solve problems, and reason with personal data in an enjoyable and immersive manner. In addition, creating \textbf{multisensory data dashboards} that combine visual representations with auditory cues, interactive touchscreens, or haptic feedback can provide a comprehensive data experience. This allows individuals to reason with personal data using a combination of visual, auditory, and tactile senses, enhancing the depth of understanding and analysis. Future visualization research may better understand how different modalities can be used to accommodate IDD individuals' complex sensory needs and preferences, and more thoroughly examine the impact of various modalities on cognitive accessibility.

\subsection{Creating Culturally Relevant Data Interfaces}
Since IDD is usually a set of disabilities rather than single diagnosis, it may impact each individual differently \cite{faqs}. In our experiences working with this population, almost everyone has a distinctive set of conditions and faces unique challenges despite being characterized under the umbrella term ``IDD,'' which makes developing a universally accessible solution extremely difficult. Yet our graphical perception experiment (Fig. \ref{fig:guide}) demonstrated that thoughtfully designed visualizations can serve as a cognitive amplifier and help 
people with a range of disabilities
better reason with data \cite{chi}. For example, people with IDD tend to relate abstract visual elements to real-world objects (e.g., treemaps as colored pieces of paper, stacked bars as rising stairs etc.). They benefit from familiar visual metaphors, such as dollar signs and other semantically meaningful imagery, to understand the context of data. Additionally, axis-aligned isotype visualizations and discrete, countable representations can also largely improve their accuracy and confidence in data interpretation. While pie charts are extremely inaccessible, stacked bars and treemaps can make proportion data twice as accessible. However, our experiment only covered a niche application of data visualization (fiscal analysis). Most conventional tools and guidelines were developed for neurotypical consumers without considering the differences of people with IDD, introducing unintentional yet ongoing challenges for this population. Cognitively accessible visualizations will require a different set of guidelines and likely take on different shapes \cite{chi}. They should be designed with respect to the unique needs, abilities, and preferences of people with IDD, tapping into their shared cultural knowledge and experiences. Designing such a culturally relevant interface may involve several aspects: 

\begin{figure}[h]
\centering
\includegraphics[width=8cm]{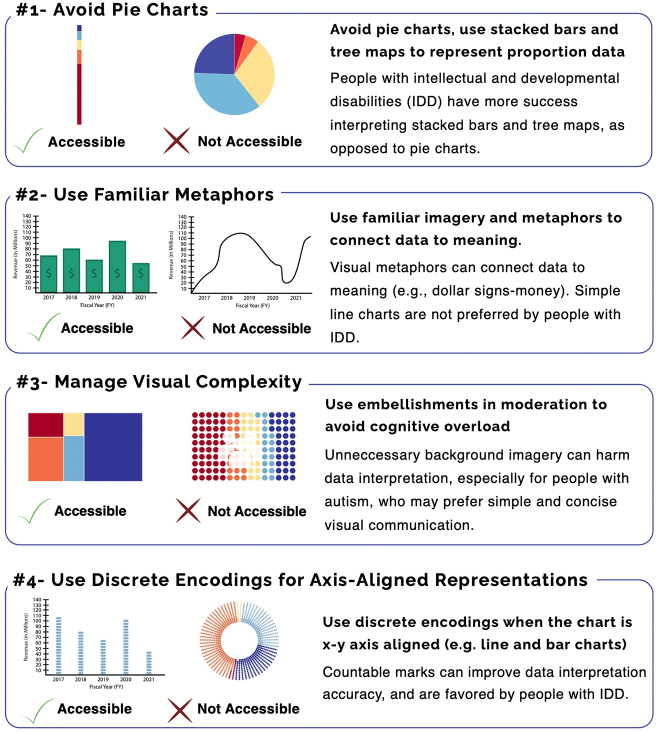}
  \caption{%
  	Guidelines for designing cognitively accessible visualizations drawing from the graphical perception experiment in support of financial self-advocacy for people with IDD.%
  }
  \label{fig:guide}
\end{figure}

\noindent\textbf{(1) Visual Design.} Incorporating culturally appropriate visuals, colors, symbols and imagery that are familiar and meaningful to people with IDD in the design. Our studies suggest that individuals with IDD generally prefer true-to-life representations over abstract visualization when reasoning with data. When designing for a particular scenario, designers may first need to understand what visuals make sense to people with IDD, and then consider maximizing the details of the representation, using photos, icons and other relevant pictorials to communicate data context. Future research should examine the role of semiotics in designing accessible visualizations or even explore what ``useful'' chart junk \cite{cj} would look like for people with IDD, identifying meaningful signs and symbols to include in visualization design to cater to their needs.

\noindent\textbf{(2) Navigation \& Layout.} Organizing the navigation and layout of a visualization to match IDD individuals' expectations and mental models. People with IDD usually have attention and memory issues \cite{faqs}, whereas those with ASD specifically are highly susceptible to cognitive overload and can easily get overwhelmed by unexpected changes and transitions \cite{sensory}. However, both groups mentioned the potential benefit of guided sensemaking, such as ``step-by-step'' imagery, to help direct their attention \cite{datachi}. Future research should explore the impact of turning abstract visualizations into more concrete formats, such as data comics \cite{comics}, data videos\cite{videos}, or games \cite{games}, as they may enable self-paced navigation and help people with IDD break down multidimensional data into digestible pieces. 

\noindent\textbf{(3) Content, Language \& Representation.} Adapting the texts and communication styles of the visualization to reflect the preferences of people with IDD. To ensure the content is inclusive and make people feel represented and respected, designers need to be careful with their data selection and use unbiased data sources, paying close attention to the visual encoding and design choices to not skew or distort the data. They need to provide contextual information and consider alternative perspectives, encouraging critical thinking and interpretation of the data. The visualization needs to be an objective, accurate, and inclusive representation of the data, and the people it describes. Future research may explore how affective design elements \cite{affect} can meaningfully reflect the cultural identities of this population and create emotionally engaging visualizations.

\section{The Need for Cognitively Diverse Visualizations} 
People with Intellectual and Developmental Disabilities (IDD) encompass a diverse population with unique characteristics and abilities, which often translate to more complex and intense challenges compared with other disabilities. The solution for developing cognitively accessible visualization is not a straightforward one, as it needs to cater to diverse cognitive needs and abilities, accounting for different information processing, comprehension and interaction preferences and challenges, and will require flexible and adaptable approaches to working with this population and presenting information. Unlike physical disabilities that can often be addressed through standardized accessibility guidelines, cognitive disabilities are highly individualized and context-dependent. There is no one-size-fits-all approach to cognitive accessibility, making it challenging to develop universal guidelines or standards. 

Yet, our preliminary investigations show that visualization could be a promising tool to improve cognitive access to data and provide concrete support for people with IDD in various areas of life. However, designing cognitively accessible visualizations will require a different set of guidelines and concerted efforts dedicated to addressing their particular challenges. We argue that designing for cognitive accessibility is to design for cognitive diversity, and it is a matter of equity and respect. By providing equitable access to information, individuals can gain knowledge, express their opinions, and contribute to decision-making processes. And by designing visualizations that cater to diverse cognitive needs, we can foster a more inclusive and equitable society where data is accessible to all.

\section{Conclusion}
Data can empower people with IDD in many ways. However, working with data is difficult and often out of the reach of this population. In this paper, we argue for a near-term research agenda for cognitively accessible visualizations to serve as an equalizer for people with IDD to better participate in society. Based on preliminary findings and our own reflections working with this population and their data, we discuss two immediate research goals for cognitively accessible visualizations to improve data accessibility. First, visualizations should support people with IDD with better personal data storytelling for more effective self-advocacy and self-expression. Second, visualizations should balance novelty and familiarity in design to increase IDD individuals' awareness of and agency over personal data and ultimately make this population more visible in the inclusive data visualization space. 

\acknowledgments{
We would like to thank our colleagues, collaborators, and participants for their support. This work was funded by NSF \#2046725, \#2320920, \#2233316, and \#1933915, as well as the Ray Hauser Award at University of Colorado Boulder. }
\bibliographystyle{abbrv-doi}

\bibliography{template}

\begin{thebibliography}{10}

\bibitem{personcentered}
Dataset on {Intellectual} and {Developmental} {Disabilities}: {Linking} {Data}
  to {Enhance} {Person}-{Centered} {Outcomes} {Research}.

\bibitem{faqs}
AAIDD.
\newblock {F}{A}{Q}s on {I}ntellectual {D}isability --- aaidd.org.
\newblock
  \url{https://www.aaidd.org/intellectual-disability/faqs-on-intellectual-disability},
  2022.

\bibitem{videos}
F.~Amini, N.~Riche, B.~Lee, C.~Hurter, and P.~Irani.
\newblock Understanding data videos: Looking at narrative visualization through
  the cinematography lens.
\newblock 04 2015.

\bibitem{sensory}
J.~Ashburner, L.~Bennett, S.~Rodger, and J.~Ziviani.
\newblock Understanding the sensory experiences of young people with autism
  spectrum disorder: A preliminary investigation.
\newblock {\em Australian occupational therapy journal}, 60:171--80, 06 2013.
  doi: {{%
10\hspace{.1pt}\discretionary{.}{%
}{.}\hspace{.4pt}1111\discretionary{/}{%
}{/}1440\discretionary{%
}{-}{-}1630\hspace{.1pt}\discretionary{.}{%
}{.}\hspace{.4pt}12025}}


\bibitem{comics}
B.~Bach, N.~Henry~Riche, S.~Carpendale, and H.~Pfister.
\newblock The emerging genre of data comics.
\newblock {\em IEEE Computer Graphics and Applications}, 38:6--13, 05 2017.
  doi: {{%
10\hspace{.1pt}\discretionary{.}{%
}{.}\hspace{.4pt}1109\discretionary{/}{%
}{/}MCG\hspace{.1pt}\discretionary{.}{%
}{.}\hspace{.4pt}2017\hspace{.1pt}\discretionary{.}{%
}{.}\hspace{.4pt}33}}


\bibitem{cj}
S.~Bateman, R.~Mandryk, C.~Gutwin, A.~Genest, D.~Mcdine, and C.~Brooks.
\newblock Useful junk? the effects of visual embellishment on comprehension and
  memorability of charts.
\newblock vol.~4, pp. 2573--2582, 04 2010. doi: {{%
10\hspace{.1pt}\discretionary{.}{%
}{.}\hspace{.4pt}1145\discretionary{/}{%
}{/}1753326\hspace{.1pt}\discretionary{.}{%
}{.}\hspace{.4pt}1753716}}


\bibitem{datasculpture}
R.~Bhargava and C.~D’Ignazio.
\newblock Data sculptures as a playful and low-tech introduction to working
  with data.
\newblock In {\em Designing Interactive Systems Conference}, DIS '17.
  Association for Computing Machinery, 2017. doi: {{%
1721\hspace{.1pt}\discretionary{.}{%
}{.}\hspace{.4pt}1\discretionary{/}{%
}{/}123453}}


\bibitem{braddock}
D.~L. Braddock, R.~Hemp, E.~S. Tanis, J.~Wu, L.~Haffer, A.~A.~O. Intellectual,
  and D.~Disabilities.
\newblock {\em The state of the states in intellectual and developmental
  disabilities, 2017}.
\newblock American Association On Intellectual And Developmental Disabilities,
  2017.

\bibitem{cbs}
J.~Brown, M.~Brown, and P.~Dibiasio.
\newblock Treating individuals with intellectual disabilities and challenging
  behaviors with adapted dialectical behavior therapy.
\newblock {\em Journal of mental health research in intellectual disabilities},
  6:280--303, 10 2013. doi: {{%
10\hspace{.1pt}\discretionary{.}{%
}{.}\hspace{.4pt}1080\discretionary{/}{%
}{/}19315864\hspace{.1pt}\discretionary{.}{%
}{.}\hspace{.4pt}2012\hspace{.1pt}\discretionary{.}{%
}{.}\hspace{.4pt}700684}}


\bibitem{physical}
S.-A. Cooper, G.~McLean, B.~Guthrie, A.~McConnachie, S.~Mercer, F.~Sullivan,
  and J.~Morrison.
\newblock Multiple physical and mental health comorbidity in adults with
  intellectual disabilities: Population-based cross-sectional analysis.
\newblock {\em BMC family practice}, 16:110, 08 2015. doi: {{%
10\hspace{.1pt}\discretionary{.}{%
}{.}\hspace{.4pt}1186\discretionary{/}{%
}{/}s12875\discretionary{%
}{-}{-}015\discretionary{%
}{-}{-}0329\discretionary{%
}{-}{-}3}}


\bibitem{DRS}
T.~K. K.~H. Dumičić, Ž. and G.~Joost.
\newblock Design elements in data physicalization: A systematic literature
  review.
\newblock jun 2022. doi: {{%
10\hspace{.1pt}\discretionary{.}{%
}{.}\hspace{.4pt}21606\discretionary{/}{%
}{/}drs\hspace{.1pt}\discretionary{.}{%
}{.}\hspace{.4pt}2022\hspace{.1pt}\discretionary{.}{%
}{.}\hspace{.4pt}660}}


\bibitem{DIgnazio2018CreativeDL}
C.~D’Ignazio and R.~Bhargava.
\newblock Creative data literacy: A constructionist approach to teaching
  information visualization.
\newblock {\em Digit. Humanit. Q.}, 12, 2018.

\bibitem{intervention}
P.~Hamers, D.~Festen, and H.~Hermans.
\newblock Non-pharmacological interventions for adults with intellectual
  disabilities and depression: a systematic review: Non-pharmacological
  interventions for depression.
\newblock {\em Journal of Intellectual Disability Research}, 62, 05 2018. doi:
  {{%
10\hspace{.1pt}\discretionary{.}{%
}{.}\hspace{.4pt}1111\discretionary{/}{%
}{/}jir\hspace{.1pt}\discretionary{.}{%
}{.}\hspace{.4pt}12502}}


\bibitem{book}
J.~Harris.
\newblock {\em Intellectual Disability: Understanding Its Development, Causes,
  Classification, Evaluation, and TreatmentUnderstanding Its Development,
  Causes, Classification, Evaluation, and Treatment}.
\newblock 11 2005. doi: {{%
10\hspace{.1pt}\discretionary{.}{%
}{.}\hspace{.4pt}1093\discretionary{/}{%
}{/}oso\discretionary{/}{%
}{/}9780195178852\hspace{.1pt}\discretionary{.}{%
}{.}\hspace{.4pt}001\hspace{.1pt}\discretionary{.}{%
}{.}\hspace{.4pt}0001}}


\bibitem{constructive}
S.~Huron, S.~Carpendale, A.~Thudt, A.~Tang, and M.~Mauerer.
\newblock Constructive visualization.
\newblock {\em Proceedings of the Conference on Designing Interactive Systems:
  Processes, Practices, Methods, and Techniques, DIS}, 06 2014. doi: {{%
10\hspace{.1pt}\discretionary{.}{%
}{.}\hspace{.4pt}1145\discretionary{/}{%
}{/}2598510\hspace{.1pt}\discretionary{.}{%
}{.}\hspace{.4pt}2598566}}


\bibitem{10.1111:cgf.14298}
N.~W. Kim, S.~C. Joyner, A.~Riegelhuth, and Y.-S. Kim.
\newblock {Accessible Visualization: Design Space, Opportunities, and
  Challenges}.
\newblock {\em Computer Graphics Forum}, 2021. doi: {{%
10\hspace{.1pt}\discretionary{.}{%
}{.}\hspace{.4pt}1111\discretionary{/}{%
}{/}cgf\hspace{.1pt}\discretionary{.}{%
}{.}\hspace{.4pt}14298}}


\bibitem{data4health}
G.~Krahn.
\newblock A call for better data on prevalence and health surveillance of
  people with intellectual and developmental disabilities.
\newblock {\em Intellectual and developmental disabilities}, 57:357--375, 10
  2019. doi: {{%
10\hspace{.1pt}\discretionary{.}{%
}{.}\hspace{.4pt}1352\discretionary{/}{%
}{/}1934\discretionary{%
}{-}{-}9556\discretionary{%
}{-}{-}57\hspace{.1pt}\discretionary{.}{%
}{.}\hspace{.4pt}5\hspace{.1pt}\discretionary{.}{%
}{.}\hspace{.4pt}357}}


\bibitem{healthproblem}
H.~Lantman and P.~Walsh.
\newblock Managing health problems in people with intellectual disabilities.
\newblock {\em BMJ (Clinical research ed.)}, 337:a2507, 02 2008. doi: {{%
10\hspace{.1pt}\discretionary{.}{%
}{.}\hspace{.4pt}1136\discretionary{/}{%
}{/}bmj\hspace{.1pt}\discretionary{.}{%
}{.}\hspace{.4pt}a2507}}


\bibitem{affect}
E.~Lee-Robbins and E.~Adar.
\newblock Affective learning objectives for communicative visualizations.
\newblock {\em IEEE Transactions on Visualization and Computer Graphics},
  29(01):1--11, jan 2023. doi: {{%
10\hspace{.1pt}\discretionary{.}{%
}{.}\hspace{.4pt}1109\discretionary{/}{%
}{/}TVCG\hspace{.1pt}\discretionary{.}{%
}{.}\hspace{.4pt}2022\hspace{.1pt}\discretionary{.}{%
}{.}\hspace{.4pt}3209500}}


\bibitem{10.1371/journal.pone.0256294}
P.~Liao, C.~Vajdic, J.~Trollor, and S.~Reppermund.
\newblock Prevalence and incidence of physical health conditions in people with
  intellectual disability – a systematic review.
\newblock {\em PLOS ONE}, 16(8):1--19, 08 2021. doi: {{%
10\hspace{.1pt}\discretionary{.}{%
}{.}\hspace{.4pt}1371\discretionary{/}{%
}{/}journal\hspace{.1pt}\discretionary{.}{%
}{.}\hspace{.4pt}pone\hspace{.1pt}\discretionary{.}{%
}{.}\hspace{.4pt}0256294}}


\bibitem{ognjenovic1997psiholovska}
P.~Ognjenovi{\'c} and K.~Andre.
\newblock {\em Psiholo{\v{s}}ka teorija umetnosti}.
\newblock Institut za psihologiju, 1997.

\bibitem{ada}
S.~Pappas.
\newblock Despite the {ADA}, equity is still out of reach, 2020.

\bibitem{games}
A.~Rapp.
\newblock Gamification for self-tracking: From world of warcraft to the design
  of personal informatics systems.
\newblock In {\em Proceedings of the 2018 CHI Conference on Human Factors in
  Computing Systems}, CHI '18, p. 1–15. Association for Computing Machinery,
  New York, NY, USA, 2018. doi: {{%
10\hspace{.1pt}\discretionary{.}{%
}{.}\hspace{.4pt}1145\discretionary{/}{%
}{/}3173574\hspace{.1pt}\discretionary{.}{%
}{.}\hspace{.4pt}3173654}}


\bibitem{JansenvanVuuren2020StigmaAA}
J.~J. van Vuuren and H.~M. Aldersey.
\newblock Stigma, acceptance and belonging for people with idd across cultures.
\newblock {\em Current Developmental Disorders Reports}, 7:163 -- 172, 2020.

\bibitem{chi}
K.~Wu, E.~Petersen, T.~Ahmad, D.~Burlinson, S.~Tanis, and D.~A. Szafir.
\newblock Understanding data accessibility for people with intellectual and
  developmental disabilities.
\newblock In {\em Proceedings of the 2021 CHI Conference on Human Factors in
  Computing Systems}, CHI '21. Association for Computing Machinery, New York,
  NY, USA, 2021. doi: {{%
10\hspace{.1pt}\discretionary{.}{%
}{.}\hspace{.4pt}1145\discretionary{/}{%
}{/}3411764\hspace{.1pt}\discretionary{.}{%
}{.}\hspace{.4pt}3445743}}


\bibitem{datachi}
K.~Wu, M.~H. Tran, E.~Petersen, V.~Koushik, and D.~A. Szafir.
\newblock Data, data, everywhere: Uncovering everyday data experiences for
  people with intellectual and developmental disabilities.
\newblock In {\em Proceedings of the 2023 CHI Conference on Human Factors in
  Computing Systems}, CHI '23. Association for Computing Machinery, New York,
  NY, USA, 2023. doi: {{%
10\hspace{.1pt}\discretionary{.}{%
}{.}\hspace{.4pt}1145\discretionary{/}{%
}{/}3544548\hspace{.1pt}\discretionary{.}{%
}{.}\hspace{.4pt}3581204}}


\bibitem{mentalhealth}
Y.~Zisman-Ilani.
\newblock The mental health crisis of individuals with intellectual and
  developmental disabilities.
\newblock {\em Psychiatric Services}, 73:245--246, 03 2022. doi: {{%
10\hspace{.1pt}\discretionary{.}{%
}{.}\hspace{.4pt}1176\discretionary{/}{%
}{/}appi\hspace{.1pt}\discretionary{.}{%
}{.}\hspace{.4pt}ps\hspace{.1pt}\discretionary{.}{%
}{.}\hspace{.4pt}202200022}}


\end{thebibliography}
\end{document}